\documentclass[12pt]{article}
\usepackage{lscape}
\usepackage{citesort}
\usepackage{amssymb}
\usepackage{appendix}
\usepackage{multirow}

\begin{document}

\def\qq{\langle \bar q q \rangle}
\def\uu{\langle \bar u u \rangle}
\def\dd{\langle \bar d d \rangle}
\def\sp{\langle \bar s s \rangle}
\def\GG{\langle g_s^2 G^2 \rangle}
\def\Tr{\mbox{Tr}}
\def\figt#1#2#3{
        \begin{figure}
        $\left. \right.$
        \vspace*{-2cm}
        \begin{center}
        \includegraphics[width=10cm]{#1}
        \end{center}
        \vspace*{-0.2cm}
        \caption{#3}
        \label{#2}
        \end{figure}
	}
	
\def\figb#1#2#3{
        \begin{figure}
        $\left. \right.$
        \vspace*{-1cm}
        \begin{center}
        \includegraphics[width=10cm]{#1}
        \end{center}
        \vspace*{-0.2cm}
        \caption{#3}
        \label{#2}
        \end{figure}
                }

\def\ds{\displaystyle}
\def\beq{\begin{equation}}
\def\eeq{\end{equation}}
\def\bea{\begin{eqnarray}}
\def\eea{\end{eqnarray}}
\def\beeq{\begin{eqnarray}}
\def\eeeq{\end{eqnarray}}
\def\ve{\vert}
\def\vel{\left|}
\def\ver{\right|}
\def\nnb{\nonumber}
\def\ga{\left(}
\def\dr{\right)}
\def\aga{\left\{}
\def\adr{\right\}}
\def\lla{\left<}
\def\rra{\right>}
\def\rar{\rightarrow}
\def\lrar{\leftrightarrow}  
\def\nnb{\nonumber}
\def\la{\langle}
\def\ra{\rangle}
\def\ba{\begin{array}}
\def\ea{\end{array}}
\def\tr{\mbox{Tr}}
\def\ssp{{\Sigma^{*+}}}
\def\sso{{\Sigma^{*0}}}
\def\ssm{{\Sigma^{*-}}}
\def\xis0{{\Xi^{*0}}}
\def\xism{{\Xi^{*-}}}
\def\qs{\la \bar s s \ra}
\def\qu{\la \bar u u \ra}
\def\qd{\la \bar d d \ra}
\def\qq{\la \bar q q \ra}
\def\gGgG{\la g^2 G^2 \ra}
\def\GG{\langle g_s^2 G^2 \rangle}
\def\g5{\gamma_5 \not\!q}
\def\x{\gamma_5 \not\!x}
\def\g5{\gamma_5}
\def\sb{S_Q^{cf}}
\def\sd{S_d^{be}}
\def\su{S_u^{ad}}
\def\sbp{{S}_Q^{'cf}}
\def\sdp{{S}_d^{'be}}
\def\sup{{S}_u^{'ad}}
\def\ssp{{S}_s^{'??}}

\def\sig{\sigma_{\mu \nu} \gamma_5 p^\mu q^\nu}
\def\fo{f_0(\frac{s_0}{M^2})}
\def\ffi{f_1(\frac{s_0}{M^2})}
\def\fii{f_2(\frac{s_0}{M^2})}
\def\O{{\cal O}}
\def\sl{{\Sigma^0 \Lambda}}
\def\es{\!\!\! &=& \!\!\!}
\def\ap{\!\!\! &\approx& \!\!\!}
\def\ar{&+& \!\!\!}
\def\arrr{\!\!\!\! &+& \!\!\!}
\def\ek{&-& \!\!\!}
\def\vev{&\vert& \!\!\!}
\def\kek{\!\!\!\!&-& \!\!\!}
\def\cp{&\times& \!\!\!}
\def\se{\!\!\! &\simeq& \!\!\!}
\def\eqv{&\equiv& \!\!\!}
\def\kpm{&\pm& \!\!\!}
\def\kmp{&\mp& \!\!\!}
\def\mcdot{\!\cdot\!}
\def\erar{&\rightarrow&}


\def\simlt{\stackrel{<}{{}_\sim}}
\def\simgt{\stackrel{>}{{}_\sim}}


\title{
         {\Large
                 {\bf
Analysis of the radiative decays $\Sigma_Q \to \Lambda_Q \gamma$ and
$\Xi^\prime_Q \to \Xi_Q \gamma$ in  light cone sum rules
                 }
         }
      }

\author{\vspace{1cm}\\
{\small T. M. Aliev$^{1,}$\thanks {e-mail:
taliev@metu.edu.tr}~\footnote{permanent address:Institute of
Physics,Baku,Azerbaijan}\,\,,
T. Barakat$^{2,}$\thanks {e-mail:
tbarakat@KSU.EDU.SA}\,\,,
M. Savc{\i}$^{1,}$\thanks
{e-mail: savci@metu.edu.tr}} \\
{\small $^1$ Physics Department, Middle East Technical University,
06800 Ankara, Turkey }\\
{\small $^2$ Physics and Astronomy Department, King Saud University, Saudi Arabia}}

\date{}

\begin{titlepage}
\maketitle
\thispagestyle{empty}

\begin{abstract}

The light cone sum rules method is used in studying the radiative decays
$\Sigma_Q \to \Lambda_Q \gamma$ and $\Xi^\prime_Q \to \Xi_Q \gamma$.
Firstly, the sum rules for the form factor $F_2(Q^2=0)$ responsible for
these transitions is constructed. Using this result the decay widths of the
above--mentioned decays are calculated and analyzed. A comparison of our
predictions on the decay widths of considered transitions with the
predictions of the other approaches is presented.

\end{abstract}

~~~PACS numbers: 11.55.Hx, 13.40.Hq, 14.40.Lb, 14.40.Nd

\end{titlepage}

\section{Introduction}

In the past decade revolutionary progress has been made in hadron spectroscopy.
Many baryons with single heavy quark have been observed experimentally
\cite{Rmrt01}. At the same time many new charmonium and bottomonium like
states have also been discovered \cite{Rmrt02}. These states have more
complicated structure compared to the ones predicted by the quark model.

These experimental achievements have carried the studies to a new level,
namely, the study of the decays of these baryons. In this work we
concentrate our attention on the heavy baryons with single heavy quark, and
investigate their electromagnetic decays.

According to $SU(3)$ classification, the heavy baryons ground states with
spin--1/2 belong to the sextet representation; and with spin--3/2 to the
sextet and spin--1/2 to the anti--triplet representations.
It is customary to denote these representations as $6$, $6^\ast$ and
$\bar{3}$.

The radiative decays among the baryons belonging to these representations
have already been studied in framework of the
nonrelativistic potential model \cite{Rmrt03}, light cone QCD sum rules
incorporating with the heavy quark effective theory \cite{Rmrt04}, and
incorporating both with heavy and chiral symmetry \cite{Rmrt05};
$(2+1)$ flavor lattice QCD \cite{Rmrt06}; heavy hadron chiral perturbation
theory \cite{Rmrt07,Rmrt08}; chiral perturbation theory \cite{Rmrt09};
relativistic three--quark model \cite{Rmrt10}, heavy quark symmetry
\cite{Rmrt11}, static quark model \cite{Rmrt12}, bag model \cite{Rmrt13},
which lead to quite different results. Therefore further independent calculations
on these decay widths are necessary.

The present work is devoted to the study of the $\Sigma_Q \to \Lambda_Q
\gamma$ and $\Xi^\prime_Q \to \Xi_Q \gamma$ decays in framework of
the light cone QCD sum rules.  Note that, the decay widths between the
$6^\ast \to 6$ and $6^\ast \to \bar{3}$ transitions have been studied
earlier in framework of the light cone QCD sum rules method in
\cite{Rmrt14,Rmrt15}.

The paper is organized as follows. In section 2, light cone QCD sum rules for the
electromagnetic form factor $F_2(q^2=0)$ responsible for these decays is 
derived. The last section is devoted to the
numerical analysis. In this section we also present a comparison of our
predictions with the results of other approaches.

\section{Light cone QCD sum rules for the $\Sigma_Q \to \Lambda_Q \gamma$
and $\Xi_Q^\prime \to \Xi_Q \gamma$ decay form factors}   

In this section we derive the light cone QCD sum rules for the radiative
$\Sigma_Q \to \Lambda_Q \gamma$ and $\Xi_Q^\prime \to \Xi_Q \gamma$ decay
form factors. For this purpose we start with the definition of the
transition matrix element between heavy baryon states in presence of the
electromagnetic field, i.e., $\lla B_{Q_2}(p,s^\prime) \vel j_\mu^{el} \ver
B_{Q_1}(p+q,s) \rra$. This matrix element is parametrized in terms of the
Dirac $F_1(Q^2)$ and Pauli $F_2(Q^2)$ form factors in the following way:
\bea
\label{emrt01}
\lla B_{Q_2}(p,s^\prime) \vel j_\mu^{el} \ver  B_{Q_1}(p+q,s) \rra \es
\bar{u}(p) \Bigg[ \left(\gamma_\mu - {\not\!{q} q_\mu \over q^2} \right)
F_1(Q^2) \nnb \\
\ek {1 \over m_{B_{Q_1}} + m_{B_{Q_2}}}
i \sigma_{\mu\nu} q^\nu F_2(Q^2) \Bigg] u(p+q)~.
\eea 
For the real photons, obviously, we need to know the values of these form
factors only at the point $Q^2=-q^2=0$. This process, i.e., transition of
one of the heavy baryon in the sextet representation, to another heavy
baryon in the anti--triplet representation in presence of the
electromagnetic field, is described by the following correlation function,
\bea
\label{emrt02}
\Pi_\mu (p,q) \es - \int d^4x \int d^4y e^{i(px+qy)} \lla 0 \vel T\left\{
\eta_{Q_1}^a (0) \, j_\mu (y) \,\bar{\eta}_{Q_2}^s (x) \right\} \ver 0
\rra~,
\eea
where $j_\mu = e_q \bar{q} \gamma_\mu q + e_Q \bar{Q} \gamma_\mu Q$ is the
electromagnetic current with the electric charges $e_q$ and $e_Q$ for the
light and heavy quarks; and ${\eta}_{Q_2}^s$ and ${\eta}_{Q_2}^a$
are the interpolating currents in the sextet and anti--triplet representations,
respectively.

The general form of the interpolating currents of the spin--1/2 heavy
baryons in the sextet and anti--triplet representations are given as (see
for example \cite{Rmrt16}),
\bea
\label{emrt03}
\eta_Q^s \es -{1\over \sqrt{2}} \epsilon^{abc} \Big\{ (q_1^{aT} C
Q^b) \gamma_5 q_2^c - (Q^{aT} C q_2^b) \gamma_5 q_1^c + \beta
(q_1^{aT} C \gamma_5 Q^b) q_2^c - \beta (Q^{aT} C \gamma_5 q_2^b) q_1^c
\Big\}~, \nnb \\
\eta_Q^a \es {1\over \sqrt{6}} \epsilon^{abc} \Big\{ 2 (q_1^{aT} C   
q_2^b) \gamma_5 Q^c + (q_1^{aT} C Q^b) \gamma_5 q_2^c +
(Q^{aT} C q_2^b) \gamma_5 q_1^c \nnb \\
\ar 2 \beta (q_1^{aT} C \gamma_5 q_2^b) Q^c +
(q_1^{aT} C \gamma_5 Q^b) q_2^c +
(Q^{aT} C \gamma_5 q_2^b) q_1^c \Big\}~,
\eea
where $\beta$ is the arbitrary auxiliary parameter, and the light quark
contents of the
heavy baryons in sextet and antitriplet representations are summarized in
Table 1.


\begin{table}[h]

\renewcommand{\arraystretch}{1.3}
\addtolength{\arraycolsep}{-0.5pt}
\small
$$
\begin{array}{|c|c|c|c|c|c|c|c|c|c|}
\hline \hline   
 & \Sigma_{b(c)}^{+(++)} & \Sigma_{b(c)}^{0(+)} & \Sigma_{b(c)}^{-(0)} &
             \Xi_{b(c)}^{\prime -(0)}  & \Xi_{b(c)}^{\prime 0(+)}  &
 \Lambda_{b(c)}^{0(+)} & 
\Xi_{b(c)}^{-(0)}  & \Xi_{b(c)}^{0(+)} \\   
\hline \hline
q_1   & u & u & d & d & u & u & d & u \\
q_2   & u & d & d & s & s & d & s & s \\
\hline \hline 
\end{array}
$$
\caption{Light quark contents of the heavy baryons in the symmetric sextet, and
anti--symmetric anti--triplet representations.}
\renewcommand{\arraystretch}{1}
\addtolength{\arraycolsep}{-1.0pt}
\end{table}      


Introducing a plane wave electromagnetic background field $F_{\mu\nu} = i
(\varepsilon_\mu q_\nu - \varepsilon_\nu q_\mu) e^{iqx}$, it is possible to
rewrite the correlator (\ref{emrt02}) in the following way,
\bea
\label{emrt04} 
\Pi_\mu(p,q) \varepsilon^\mu = i \int d^4x e^{ipx} \lla 0 \vel T
\left\{\eta_{Q_1}(0) \, \bar{\eta}_{Q_2} (x) \right\} \ver 0 \rra_F~,
\eea
where the subscript $F$ means that all vacuum expectation values should be evaluated in
the background field $F_{\mu\nu}$.
The correlation function given in Eq. (\ref{emrt02}) can be obtained from
Eq. (\ref{emrt03}) by expanding it in powers of the background field, and
considering only the linear term in $F_{\mu\nu}$ which corresponds to the
single photon emission. More about the details of the application of the
background field method can be found in \cite{Rmrt17} and \cite{Rmrt18}.

In order to obtain the sum rules for the electromagnetic form factors
describing the $\Sigma_Q \to \Lambda_Q \gamma$
and $\Xi_Q^\prime \to \Xi_Q \gamma$ transition the correlation function is
calculated in terms of hadrons from one side, and in terms of the
quark--gluon degrees of freedom by using the operator product expansion
(OPE) and introducing the photon distribution amplitudes (DAs) from the
other side. The photon DAs are the main nonperturbative ingredient of the
light cone sum rules. In this version of the light cone QCD sum rules OPE is
performed by twist of the nonlocal operator, rather than dimensions of the
operators in the traditional sum rules. The sum rules is obtained by
matching these two representations.

We start our analysis by constructing the correlation function from the
hadronic side. It can be obtained by inserting all intermediate hadronic sum
rules, having the same quantum numbers as the corresponding interpolating
currents $\eta_Q$. After isolating the ground state's contribution ve get,
\bea
\label{emrt05}
\Pi_\mu(p,q) \es
{\la 0 \ve \eta_{Q_2}^a \ve B_{Q_2}(p_2) \ra \over p_2^2-m_{B_{Q_2}}^2} \la
B_{Q_2}(p_2)\ve j_\mu^{el}(q) \ve
B_{Q_1}(p_1) \ra {\la B_{Q_1}(p_1) \ve \bar{\eta}_{Q_1}^s \ve 0\ra \over
p_1^2-m_{B_{Q_1}}^2} + \cdots~,
\eea
where the contributions coming from the higher states and continuum is
denoted by dots, and $p_1=p_2+q$.

The expression for the correlator function can be obtained from the
hadronic side substituting the matrix elements appearing in
Eq. (\ref{emrt02}). These matrix elements are defined in the standard way as
follows,
\bea
\label{emrt06}
\la 0 \ve \eta_{B_{Q_2}} \ve B_{Q_2}(p_2)\ra \es  \lambda_2
u_{B_{Q_2}}(p_2)~,\nnb \\
\la  B_{Q_1}(p_1) \ve \eta_{B_{Q_1}} \ve 0\ra \es  \lambda_1
\bar{u}_{B_{Q_1}}(p_1)~,\nnb \\
\la B_{Q_2}(p_2) \ve j_\mu^{el}(q) \ve B_{Q_1}(p_1)\ra \es
\bar{u}_{B_{Q_2}}(p_2)
\left[\left(\gamma_\mu - {\not\!{q} q_\mu \over q^2} \right)
F_1 - {i \sigma_{\mu\nu} q^\nu \over  m_{B_{Q_1}} 
+ m_{B_{Q_1}} } F_2 \right] u_{B_{Q_1}}(p_1)~,
\eea
where $\lambda_i$ are the residues of the hadrons, $B_{Q_i}$ are the baryons
and $m_{B_{Q_i}}$ are their respective masses, $F_1$ and $F_2$ are the Dirac
and Pauli form factors, respectively.
Using the equation of motion, the matrix element
$\la B_{Q_2}(p_2) \ve j_\mu^{el}(q) \ve B_{Q_1}(p_1)\ra$ can be written in
the following way,
\bea
\label{emrt07}
\la B_{Q_2}(p_2) \ve j_\mu^{el}(q) \ve B_{Q_1}(p_1)\ra \es
\bar{u}_{B_{Q_2}}(p_2)
\Bigg[\gamma_\mu (F_1+F_2) - {\not\!{q} q_\mu \over q^2} F_1 \nnb \\
\ek {(p_1+p_2)_\mu \over  m_{B_{Q_1}}
+ m_{B_{Q_1}} } F_2 \Bigg] u_{B_{Q_1}}(p_1)~.
\eea
Inserting Eqs. (\ref{emrt06}) and (\ref{emrt07}) into Eq. (\ref{emrt05}),
and performing summation over spins of
the Dirac spinors we get,
\bea
\label{emrt08}
\varepsilon^\mu \Pi_\mu(p,q) \es 
{ \lambda_{B_{Q_1}} \lambda_{B_{Q_2}} \over
(p^2 - m_{B_{Q_2}}^2) (p^2 - m_{B_{Q_1}}^2)} (\not\!{p} + m_{B_{Q_2}})  
\Bigg[\not\!{\varepsilon} (F_1+F_2) \nnb \\
\ek {2 (p \varepsilon) \over 
m_{B_{Q_1}} + m_{B_{Q_2}} } F_2 \Bigg] (\not\!{p} + \not\!{q} +
m_{B_{Q_1}})~, 
\eea
where we set $p_2=p$,$p_1=p+q$ and $q\varepsilon=0$.
It can easily be seen from Eq. (\ref{emrt08}) that the correlation function
possesses many structures any of them can be used for constructing the sum
rules for the form factors $F_1+F_2$ and $F_2$. The experience in working
with the sum rules shows that the structures containing maximum number of
momenta exhibit rather good convergence. For this reason in calculation of
the form factors $F_1 + F_2$ and $F_2$, we shall chose the
structures $\rlap/{p}\rlap/{\varepsilon}\rlap/{q}$ and
$\rlap/{p}(p\varepsilon)$, respectively. In this work we calculate only the form
factor $F_2$ since the transitions under consideration is described only by the form
factor $F_2$. Note that the form factor $F_1+F_2$ has already been calculated
for the transitions under consideration in \cite{Rmrt19} and \cite{Rmrt20}.   
The expression of the correlator function given in Eq. (\ref{emrt03}) can be
obtained in the deep Euclidean region 
in terms of photon DAs with increasing twist, where $p \ll 0$, and $(p+q)^2
\ll 0$.

Calculation of the correlation function can be carried out straightforwardly
using the Wick's theorem. In performing this calculation the expressions of
the light and heavy quark propagators in presence of the external field are
needed. The light quark propagator in the background field is calculated in
\cite{Rmrt21}, and it is found that the contributions of the nonlocal operators
$\bar{q} G q$, $\bar{q} G^2 q$, $\bar{q} q \bar{q} q$ are quite small.
Neglecting these contributions the expression of the light quark propagator
can be written as,
\bea
\label{emrt09}
S_q(x) \es {i \rlap/x\over 2\pi^2 x^4} - {m_q\over 4 \pi^2 x^2} -
{\lla \bar q q \rra\over 12} \left(1 - i {m_q\over 4} \rlap/x \right) -
{x^2\over 192} m_0^2 \lla \bar q q \rra  \left( 1 -
i {m_q\over 6}\rlap/x \right) \nnb \\
\ek i g_s \int_0^1 du \Bigg[{\rlap/x\over 16 \pi^2 x^2} G_{\mu \nu} (ux)
\sigma_{\mu \nu} - {i\over 4 \pi^2 x^2} u x^\mu G_{\mu \nu} (ux) \gamma^\nu
\nnb \\
\ek i {m_q\over 32 \pi^2} G_{\mu \nu} \sigma^{\mu
 \nu} \left( \ln {-x^2 \Lambda^2\over 4}  +
 2 \gamma_E \right) \Bigg]~,
\eea
where $\gamma_E$ is the Euler constant, and $\Lambda$ is the cut-off energy
separating perturbative and nonperturbative regions whose value is
calculated in \cite{Rmrt22} to be $\Lambda = (0.5 \pm 0.1)~GeV$.

The expression of the heavy quark propagator in the background field in $x$
representation is given as,
\bea
\label{emrt10}
S_Q(x) \es {m_Q^2 \over 4 \pi^2} \Bigg\{ {K_1(m_Q\sqrt{-x^2}) \over
\sqrt{-x^2}} +
i {\rlap/{x} \over \left(\sqrt{-x^2}\right)^2} K_2(m_Q\sqrt{-x^2}) \Bigg\} \nnb \\
\ek {g_s \over 16 \pi^2} \int_0^1 du
G_{\mu\nu}(ux) \left[ \left(\sigma^{\mu\nu} \rlap/x + \rlap/x
\sigma^{\mu\nu}\right) {K_1 (m_Q\sqrt{-x^2})\over \sqrt{-x^2}} +
2 \sigma^{\mu\nu} K_0(m_Q\sqrt{-x^2})\right]~,
\eea
where $K_i(m_Q\sqrt{-x^2})$ are the modified Bessel functions.

Having the expressions of the light and heavy quark propagators at hand,
calculation of the theoretical part of the correlation function is a
straightforward, but rather a tedious calculation.
Here at this point, one technical remark is in order. To be able to express
the vacuum expectation value $\lla 0 \ve q(x) \bar{q}(0) \ve 0 \rra_F$ in
terms of the photon DAs, the Fierz identity needs to be used. It should be
noted here that, our approach in calculating the nonperturbative
contribution to the correlation function follows the line of \cite{Rmrt23}
for the $D^\ast D \pi$ coupling with the replacement of the pion DAs by
the photon DAs.

As has already been noted, in constructing the sum rules for the form factor
$F_2(0)$ we have decided to choose the structure
$(\varepsilon\!\cdot\!p) \not\!{p}\!\!\not\!{q}$, in both representations of the
correlation function. In obtaining the final
result for the sum rule of the form factor $F_2(0)$, Borel transformation over
the variables $p^2$ and $(p+q)^2$ is implemented using the quark--hadron duality
anzats. Following these steps of calculation, we finally get the following
sum rule for the form factor $F_2(0)$,
\bea
\label{emrt11}
\lambda_{B_{Q_1}} \lambda_{B_{Q_2}} e^{-\left({m_{B_{Q_1}}^2\over M_1^2} +
{m_{B_{Q_2}}^2\over M_2^2}\right)} F_2(0) + \int ds_1 ds_2 \rho^h (s_1,s_2)\,
e^{-\left({s_1\over M_1^2} + {s_2 \over M_2^2}\right)} = \Pi^{B(theor)}~,
\eea
where $\lambda_{B_{Q_1}}$ and $\lambda_{B_{Q_2}}$
are the residues of the corresponding sextet and anti--triplet baryons, respectively,
whose expressions can be found in \cite{Rmrt19} and \cite{Rmrt20}; $M_1^2$
and $M_2^2$ are the Borel mass parameters for the corresponding channels.
It should be noted here that,
for consistency, the perturbative ${\cal O}(\alpha_s)$ corrections are
neglected in the calculations of residues, since they are not included in
sum rules (\ref{emrt11}). These corrections might give considerable
contribution to the form factor $F_2(0)$ similar to the $D^\ast D\pi$ case.
But, calculation
of the radiative corrections lies beyond the scope of the present work. 
Explicit expression of $ \Pi^{B(theor)}$ can be found in Appendix A. The
second term on the left hand side of Eq. (\ref{emrt11}) describes the
contributions of the higher states and continuum. In calculating the
contributions of these states we shall use the quark--hadron duality ansatz,
i.e., above some predetermined thresholds in the $(s_1,s_2)$ plane the
hadronic spectral density is replaced by the QCD spectral density
$\rho^{QCD} (s_1,s_2)$. Using this ansatz, the continuum subtraction can be
carried out by the procedure explained in \cite{Rmrt23}. Leaving aside the
technical details, in the case $M_1^2 = M_2^2 = 2 M^2$, and $u_0=1/2$,
the subtraction procedure can be performed by using the
formula,
\bea
M^{2n} e^{-m_Q^2/M^2} \to {1\over \Gamma(n)} \int_{m_Q^2}^{s_0} ds e^{-s/M^2}
\left(s-m_Q^2 \right)^{n-1}~,~~(n \ge 1)~. \nnb
\eea

We see from the expression of $\Pi^{B(theor)}$ that the leading twist term
$\varphi_\gamma (u_0)$ is proportional to $m_b^4 M^4$, and higher twist
terms are proportional to $m_b^4 M^2$ or $m_b^2 M^2$.
Therefore, higher twist terms that are
suppressed by inverse powers of $M^2$ with respect to the leading ones
remain unaffected.
Therefore, continuum subtraction procedure is not
performed for the higher twist terms
(for more detail see \cite{Rmrt23}). It should be noted here that,
in principle, single dispersion integrals originate in the subtractions which
make the double dispersion integral finite, can enter to the spectral
density. But these terms are all killed by the double Borel transformations. 

The masses of the initial and final heavy baryons are quite close to each
other, hence we can set $M_1^2 = M_2^2 = 2 M^2$, which naturally
leads to $u_0=1/2$. In our numerical analysis we use these values of $M^2$
and $u_0$.  

At the end of this section we present the formula needed to calculate the decay rate of
transitions under consideration, whose expression is as follows,
\bea
\label{emrt12}
\Gamma (B_{Q_1} \to B_{Q_2} \gamma) =
{4 \alpha \ve \vec{q} \ve^3 \over \left( m_{B_{Q_1}} \!+ m_{B_{Q_2}}\right)^2}
\ve F_2(0) \ve^2~,
\eea
where
\bea
\ve \vec{q} \ve = {\left( m_{B_{Q_1}}^2 \! - m_{B_{Q_2}}^2\right) \over
2 m_{B_{Q_1}} }~, \nnb
\eea
is the magnitude of the photon momentum.

\section{Numerical results}

In this section we perform the numerical analysis using the sum rules for the
form factor $F_2(0)$. The values input parameters used in this calculation
are as follows: The quark condensate $\uu(\mu=1~GeV)=-(0.243)^3~GeV^3$,
$\sp \ve_{\mu=1~GeV} = 0.8 \uu\ve_{\mu=1~GeV}$, $m_0^2=(0.8\pm 0.2)~GeV^2$
which is obtained from the analysis of two--point sum rules for the light
baryons \cite{Rmrt24,Rmrt25}, and $B$, $B^\ast$ \cite{Rmrt26}, respectively,
$f_{3\gamma} = -0.0039~GeV^2$ \cite{Rmrt18}, magnetic susceptibility $\chi$
which is calculated in \cite{Rmrt27,Rmrt28,Rmrt29}, where we use
$\chi(\mu=1~GeV)=-2.85~GeV^{-2}$ in the present work.

The sum rules for the form
factor $F_2(0)$ contain also three auxiliary parameters, namely, the Borel mass
parameter $M^2$, the arbitrary parameter, and
the continuum threshold $s_0$. Obviously, any physical quantity must be
independent of the above--mentioned auxiliary parameters. Therefore, we
should find the regions of these parameters for which the form factor
$F_2(0)$ shows no sensitivity to their variation. The continuum threshold is
related to the mass of the first excited state. The energy needed to excite
particle from the ground state to the first excited state is equal to
$(\sqrt{s_0}-m)$ where $m$ is the mass of the baryon in its ground state.
Usually, $(\sqrt{s_0}-m)$ varies in the interval $0.3~GeV$ and $0.8~GeV$.
The experimental values of the heavy baryons are reproduced quite well if
the continuum threshold varies in the following regions,

\bea
\label{emrt13}
\sqrt{s_0} = \left\{ \begin{array}{c}
(3.1\pm 0.1)~GeV,~\mbox{for }\Sigma_c \to \Lambda_c~, \\
(3.2\pm 0.1)~GeV,~\mbox{for }\Xi_c^\prime \to \Xi_c~, \\
(6.6\pm 0.2)~GeV,~\mbox{for }\Sigma_b \to \Lambda_b~, \\
(6.7\pm 0.2)~GeV,~\mbox{for }\Xi_b^\prime \to \Xi_b~.
\end{array} \right.
\eea
      
The upper and lower bounds of the Borel mass parameter $M^2$ are determined
by imposing the following two conditions:\\
a) The contributions of the higher
states and continuum should be less than the contributions of the ground
state.\\
b) Contributions of the higher twist terms should be less than the
contributions of the leading twist terms.\\
As the result of these two conditions, the ``working regions" of the Borel
parameter for the transitions under consideration is determined to be,
\bea
\label{emrt14}
&&2.0~GeV^2 \le M^2 \le 3.0~GeV^2,~\mbox{for }\Sigma_c \to \Lambda_c \gamma~,\nnb  \\
&&2.2~GeV^2 \le M^2 \le 3.4~GeV^2,~\mbox{for }\Xi_c^\prime \to \Xi_c \gamma~,\nnb  \\
&&5.0~GeV^2 \le M^2 \le 7.0~GeV^2,~\mbox{for }\Sigma_b \to \Lambda_b \gamma~,\nnb  \\
&&5.0~GeV^2 \le M^2 \le 7.5~GeV^2,~\mbox{for }\Xi_b^\prime \to \Xi_b \gamma~.
\eea

In order to find the working region of the arbitrary parameter $\beta$ for
the transitions under consideration,
we have studied the dependence of $F_2(0)$ on $\cos\theta$, where
$\beta=\tan\theta$, at several fixed values
of the continuum threshold $s_0$, and
Borel parameters $M^2$ chose from the working regions given in Eqs.
(\ref{emrt13}) and (\ref{emrt14}), respectively.
Our numerical analysis shows that, in the domain $-0.7\le
\cos\theta \le -0.4$, which is common for all the considered radiative
decays, the form factor $F_2(0)$ is practically independent of the arbitrary
parameter $\beta$, and we finally obtain the following values for the form
factor $F_2(0)$,
\bea
\label{emrt18}
F_2(0) = \left\{ \begin{array}{rl}
(3.0\pm 0.5)&\mbox{for }\Sigma_c^+ \to \Lambda_c^+ \gamma~, \\
(2.5\pm 0.4)&\mbox{for }\Xi_c^{\prime +} \to \Xi_c^+ \gamma~,\\
(0.45\pm 0.05)&\mbox{for }\Xi_c^{\prime 0} \to \Xi_c^0 \gamma~,\\
(10.0\pm 2.0)&\mbox{for }\Sigma_b^0 \to \Lambda_b^0 \gamma~,\\
(9.0\pm 2.0)&\mbox{for }\Xi_b^{\prime 0} \to \Xi_b^0 \gamma~,\\
(2.4\pm 0.5)&\mbox{for }\Xi_b^{\prime -} \to \Xi_b^- \gamma~.
\end{array} \right. \nnb
\eea
Note that, exact $SU(3)$ $U$--spin flavor symmetry forbids the
$\Xi_Q^{\prime 0} \to \Xi_Q^0 \gamma$ decay. Nonzero value of $F_2(0)$ for
this decay indicates the violation of the aforementioned symmetry.

Few words about the uncertainty in determination of the form factor $F_2(0)$
are in order. The radiative ${\cal O}(\alpha_s)$ corrections can of course
bring their own uncertainty in calculation of the form factor $F_2(0)$,
which are not taken into account in the present work.
We estimate the uncertainties coming only from the errors in values of
the input parameters entering to the sum rules. 

Having calculated the values of the form factor $F_2(0)$,
we can easily calculate the values of the considered decays widths
by using Eq. (\ref{emrt14}), whose results are
summarized in Table 2. In this table, for completeness we also present the
predictions on the decay widths calculated in other approaches, such as
nonrelativistic quark model \cite{Rmrt03}, QCD sum rules method
\cite{Rmrt04}, heavy hadron chiral perturbation theory (\cite{Rmrt05},
\cite{Rmrt07} and \cite{Rmrt08}), relativistic quark model \cite{Rmrt10},
heavy quark symmetry that is implemented with the light quark symmetry
\cite{Rmrt11}, naive static quark model \cite{Rmrt12}, and bag model
\cite{Rmrt13}.



\begin{table}[!htb]

\renewcommand{\arraystretch}{1.3}
\addtolength{\arraycolsep}{-0.5pt}
\small
$$
\begin{array}{|l|c|c|c|c|c|c|c|c|c|c|}
\hline \hline 
                &\mbox{This work}& \cite{Rmrt03} & \cite{Rmrt04} 
                                 & \cite{Rmrt05} & \cite{Rmrt07} &
                   \cite{Rmrt08} & \cite{Rmrt10} & \cite{Rmrt11} &
                   \cite{Rmrt12} & \cite{Rmrt13}                       \\ \hline
\Sigma_c^+ \to \Lambda_c^+ \gamma 
                 &  50.0 \pm 17.0 & 60.55          & 60             
                                  & 91.5           & \mbox{--}      &
                    164           & 60.7 \pm 1.5   & 87             &
                    120           & 46                                 \\
\Xi_c^{\prime +} \to \Xi_c^+ \gamma
                 &  8.5 \pm 2.5   & \mbox{--}      & \mbox{--}      
                                  & 19.7           & \mbox{--}      &
                    54            & 12.7 \pm 1.5   & \mbox{--}      &
                    14            & 10                                 \\
\Xi_c^{\prime 0} \to \Xi_c^0 \gamma
                 &  0.27 \pm 0.06 & \mbox{--}      & \mbox{--}             
                                  & 0.4            & 1.2 \pm 0.7    &
                    0.02          & 0.17 \pm 0.02  & \mbox{--}      &
                    0.33          & 0.0015                             \\
\Sigma_b^0 \to \Lambda_b^0 \gamma
                 &  152.0 \pm 60.0& \mbox{--}      & \mbox{--}             
                                  & \mbox{--}      & \mbox{--}      &  
                    287.65        & \mbox{--}      & \mbox{--}      &
                    \mbox{--}     & \mbox{--}                          \\
\Xi_b^{\prime 0} \to \Xi_b \gamma
                 &  47.0 \pm 21.0 & \mbox{--}      & \mbox{--}             
                                  & \mbox{--}      & \mbox{--}      &  
                    \mbox{--}     & \mbox{--}      & \mbox{--}      &
                    \mbox{--}     & \mbox{--}                          \\
\Xi_b^{\prime -} \to \Xi_b^- \gamma
                 &  3.3 \pm 1.3   & \mbox{--}      & \mbox{--}             
                                  & \mbox{--}      & 3.11\pm 1.8    &      
                    \mbox{--}     & \mbox{--}      & \mbox{--}      &
                    \mbox{--}     & \mbox{--}                          \\
\hline \hline
\end{array}   
$$
\caption{Decay widths of the $\Sigma_Q \to \Lambda_Q \gamma$ and
$\Xi_Q^\prime \to \Xi_Q \gamma$ transitions (in units of KeV)}
\renewcommand{\arraystretch}{1}
\addtolength{\arraycolsep}{-1.0pt}
\end{table}       

From the comparison of our results with those existing in literature, we see
that our predictions are closer to the predictions of the relativistic quark
model, and especially our result for the $\Xi_b^{\prime -} \to \Xi_b^- \gamma$
transition coincides with the result of \cite{Rmrt07}. We also observe that
there appears considerable difference among our results and the predictions of
the other approaches on the decay widths of the
considered transitions. Of course, only the experimental measurements of
these decays can play the ``judge" for choosing the right ``theory".

In conclusion, we calculate the form factor $F_2(0)$ for the $\Sigma_Q \to
\Lambda_Q \gamma$ and $\Xi_Q^\prime \to \Xi_Q \gamma$ transitions within
the light cone QCD sum rules method. The corresponding decay widths are
estimated by using these values of the form factor $F_2(0)$. Comparison of
our predictions on decay widths with the results of other approaches is
presented.


\newpage

\section*{Appendix}
\setcounter{equation}{0}
\setcounter{section}{0}


In this Appendix we present the explicit form of the 
the correlation function
$\Pi^{B(theor)}$ for the form factor $F_2(0)$ which is determined from the
coefficient of the 
$(\varepsilon\!\cdot\!p) \not\!{p}\!\!\not\!{q}$ structure.

%
%
%
%
\bea
\Pi^{B(theor)} \es
 {\sqrt{3} \over 128 \pi^4}
(1 - \beta^2) (e_s - e_u) m_b^3 M^4
    ({\cal I}_2 - 2 m_b^2 {\cal I}_3 + m_b^4 {\cal I}_4) \nnb \\
\ar {1\over 16 \sqrt{3} \pi^2}
  (1 - \beta)^2  \chi m_b^4 M^4 (e_s \sp - e_u \uu) ({\cal I}_3 - m_b^2 {\cal I}_4) 
   \varphi_\gamma(u_0) \nnb \\
%
%
\ar {1\over 1536 \sqrt{3} \pi^4} 
 (1 - \beta) m_b M^2 \Big\{ (1 + \beta) (e_s - e_u) \GG (3 {\cal I}_2 -
4 m_b^2 {\cal I}_3 )
+ 64 \beta  \pi^2 m_b^3 (e_u \sp - e_s \uu) {\cal I}_3 \nnb \\
%
%
\ek 24 (1 - \beta) m_b^3 \pi^2 (e_s \sp - e_u \uu) \mathbb{A} (u_0) {\cal I}_3 - 
   64 e_b m_b \pi^2 (\sp - \uu)  
    ({\cal I}_2 - m_b^2 {\cal I}_3)\Big\} \nnb \\
\ar {1\over 64 \sqrt{3} \pi^2}
(1 - \beta) m_b^2 M^2 (e_s \sp - e_u \uu) \Big\{
 \Big[(5 + \beta) {\cal I}_2 - 
    4 (2 + \beta) m_b^2 {\cal I}_3\Big] \Big[i_2({\cal S},1) 
- i_2({\cal T}_4,1) \Big]\nnb \\
\ar \Big[(1 + 5 \beta) {\cal I}_2 - 
    4 (1 + 2 \beta) m_b^2 {\cal I}_3\Big] 
\Big[i_2(\widetilde{\cal S},1) + i_2({\cal T}_2,1)\Big] 
- 8 (2 + \beta) m_b^2 {\cal I}_3 \widetilde{j}_2(h_\gamma) \nnb \\
\ar  2 (1-\beta) \Big[({\cal I}_2 - 2 m_b^2 {\cal I}_3) 
  i_2({\cal T}_1,1) + {\cal I}_2 i_2({\cal T}_3,1)\Big] \nnb \\
\ek 2 \Big[(3 + \beta) i_2({\cal S},v) + (1 + 3 \beta)
i_2(\widetilde{\cal S},v)\Big] ({\cal I}_2 - 2 m_b^2 {\cal I}_3) \nnb \\
\ek 4 \Big[\beta {\cal I}_2 - 
    (1 + \beta) m_b^2 {\cal I}_3\Big] i_2({\cal T}_2,v) -
    4 (1-\beta) {\cal I}_2 i_2({\cal T}_3,v) +
    4 \Big[{\cal I}_2 - (1 + \beta) m_b^2 {\cal I}_3\Big] 
  i_2({\cal T}_4,v \Big\}\nnb \\
\ar {1\over 32 \sqrt{3} \pi^2} 
(1 - \beta) m_b^3 M^2 f_{3\gamma} (e_s - e_u) \Big[ 2 (3 + \beta) ({\cal I}_2 -
m_b^2 {\cal I}_3) \widetilde{j}_1(\psi^v) - (1+\beta) ({\cal I}_2 - m_b^2
{\cal I}_3)  \psi^a(u_0)\Big] \nnb \\
%
%
\ar {e^{-m_b^2/M^2} \over 4608 \sqrt{3} \pi^2 M^2}
(1 - \beta) \GG (e_s \sp - e_u \uu) \Big\{3 (1 + \beta) i_2({\cal S},1) + 
     3 (1 + \beta) i_2(\widetilde{\cal S},1) \nnb \\
\ar 2 i_2({\cal T}_1,1) + 3 i_2({\cal T}_2,1) - 2 i_2({\cal T}_3,1) - 
     3 i_2({\cal T}_4,1) - 6 i_2({\cal S},v) - 2 i_2(\widetilde{\cal S},v) - 4 i_2({\cal T}_2,v) \nnb \\
\ar 4 i_2({\cal T}_3,v) + 16 \widetilde{j}_2(h_\gamma) - 
     \beta \Big[2 i_2({\cal T}_1,1) - 3 i_2({\cal T}_2,1) - 2 i_2({\cal T}_3,1) + 3 i_2({\cal T}_4,1) \nnb \\ 
\ar 2 i_2({\cal S},v) + 6 i_2(\widetilde{\cal S},v) + 4 i_2({\cal T}_3,v) - 
       4 i_2({\cal T}_4,v) - 8 \widetilde{j}_2(h_\gamma)\Big]\Big\} \nnb \\
\ar {e^{-m_b^2/M^2} \over 4608 \sqrt{3} \pi^2 M^2} 
(1 - \beta) \Big\{(1 - \beta) \GG (e_s \sp - e_u \uu) \mathbb{A} (u_0) \nnb \\ 
\ek 2 (e_u \sp - e_s \uu) \Big[3 (1 + \beta) \GG - (3 + \beta) \GG \nnb \\
\ek 8 (11 + 5 \beta) f_{3\gamma} m_0^2 \pi^2 \widetilde{j}_1(\psi^v) - 
        4 (2 + 5 \beta) f_{3\gamma} m_0^2 \pi^2 \psi^a(u_0) \Big]\Big\} \nnb \\
%
%
\ar {e^{-m_b^2/M^2} \over 96 \sqrt{3} M^4} 
(1 - \beta) f_{3\gamma} m_0^2 m_b^2 (e_u \sp - e_s \uu) 
   \Big[2 (3 + \beta) \widetilde{j}_1(\psi^v) + \beta \psi^a(u_0)\Big] \nnb \\
%
%
\ar {e^{-m_b^2/M^2} \over 13824 \sqrt{3} \pi^2 M^6 }
 (1 - \beta) \GG m_b^2 (e_u \sp - e_s \uu) 
   \Big\{3 \beta m_0^2
+ 8 f_{3\gamma} \pi^2 \Big[2 (3 + \beta) \widetilde{j}_1(\psi^v) + \beta
\psi^a(u_0)\Big]\Big\} \nnb \\
%
%
\ar {e^{-m_b^2/M^2} \over 3456 \sqrt{3} M^8} 
(1 - \beta) f_{3\gamma} \GG m_0^2 m_b^2 (e_u \sp - e_s \uu) 
    \Big[ 2 (3 + \beta) \widetilde{j}_1(\psi^v) + 
\beta \psi^a(u_0) \Big]\nnb \\
%
%
\ek {e^{-m_b^2/M^2} \over 6912 \sqrt{3} M^{10}}
(1 - \beta) f_{3\gamma} \GG m_0^2 m_b^4 (e_u \sp - e_s \uu) 
   \Big[ 2 (3 + \beta) \widetilde{j}_1(\psi^v) +
\beta \psi^a(u_0) \Big]\nnb \\
%
%
\ek {e^{-m_b^2/M^2} \over 64 \sqrt{3} \pi^2}
(1 - \beta) \beta  m_0^2 (e_u \sp - e_s \uu)  \nnb \\
\ek {1 \over 384 \sqrt{3} \pi^2} (1 - \beta) (2 - \beta) m_0^2 m_b^2  
\Big[ (e_u \sp - e_s \uu)
+ e_b (\sp - \uu) \Big] {\cal I}_2
\nnb \\
%
%
\ek {e^{-m_b^2/M^2} \over 1152 \sqrt{3} \pi^2 m_b}
(1 - \beta) (3 + \beta) f_{3\gamma} \Big[(e_s - e_u) \GG \nnb \\
\ar  96 m_b \pi^2 (e_u \sp - e_s \uu) - 3 m_b^2 e^{m_b^2/M^2} (e_s - e_u) \GG
     {\cal I}_2 \Big] \widetilde{j}_1(\psi^v) \nnb \\
%
%
\ek {1 \over 1152 \sqrt{3} \pi^2}
(1 - \beta)^2 \GG m_b^2 (e_s \sp - e_u \uu) \chi {\cal I}_2 \varphi_\gamma(u_0) \nnb \\
%
%
\ek {e^{-m_b^2/M^2} \over 2304 \sqrt{3}}
(1 - \beta) f_{3\gamma} \Bigg\{96 \beta (e_u \sp - e_s \uu) \nnb \\
\ek {1\over \pi^2 m_b} (1 + \beta) (e_s - e_u) \GG (1 - 3 m_b^2 e^{m_b^2/M^2} {\cal I}_2)
\Bigg\} \psi^a(u_0)~. \nnb
\eea
The functions $i_\ell(\phi,f(v))$, $\widetilde{j}_\ell(f(u))$, where
$(\ell=1,2)$; and
${\cal I}_n$ entering into the correlation function $\Pi^{B(theor)}$ 
are defined as:
\bea
\label{nolabel}
i_1(\phi,f(v)) \es \int {\cal D}\alpha_i \int_0^1 dv 
\phi(\alpha_{\bar{q}},\alpha_q,\alpha_g) f(v) \delta^\prime(k-u_0)~, \nnb \\
i_2(\phi,f(v)) \es \int {\cal D}\alpha_i \int_0^1 dv 
\phi(\alpha_{\bar{q}},\alpha_q,\alpha_g) f(v) \delta^{\prime\prime}(k-u_0)~, \nnb \\
\widetilde{j}_1(f(u)) \es \int_{u_0}^1 du f(u)~, \nnb \\
\widetilde{j}_2(f(u)) \es \int_{u_0}^1 du (u-u_0) f(u)~, \nnb \\
{\cal I}_n \es \int_{m_b^2}^{\infty} ds\, {e^{-s/M^2} \over s^n}~,\nnb
\eea
where 
\bea
k = \alpha_q + \alpha_g \bar{v}~,~~~~~u_0={M_1^2 \over M_1^2
+M_2^2}~,~~~~~M^2={M_1^2 M_2^2 \over M_1^2 +M_2^2}~.\nnb
\eea


\end{document}